\documentclass[fleqn,usenatbib]{mnras}

\usepackage{newtxtext,newtxmath}

\usepackage[T1]{fontenc}

\DeclareRobustCommand{\VAN}[3]{#2}
\let\VANthebibliography\thebibliography
\def\thebibliography{\DeclareRobustCommand{\VAN}[3]{##3}\VANthebibliography}

\usepackage{graphicx}	
\usepackage{amsmath}	

\def\m2002 {[M2002] SMC 12102}
\newcommand{\bexrb}{BeXRB}
\newcommand{\bexrbs}{BeXRBs}
\newcommand\Van[1]{#1}

\title[SXP 15.6]{SXP 15.6 - an accreting pulsar close to spin equilibrium?}

\author[M. J. Coe et al.]{M.~J. Coe$^{1}$\thanks{E-mail: mjcoe@soton.ac.uk},
  I. M. Monageng$^{2,4}$,  J.~A. Kennea$^{3}$, D.A.H. Buckley $^{2}$, P.~A. Evans$^{5}$, A. Udalski$^{6}$, 
  \newauthor
  Paul Groot$^{2,7,8}$, Steven Bloemen$^{7}$, 
Paul Vreeswijk$^{7}$, Vanessa McBride$^{9}$, Marc Klein-Wolt$^{7}$, \newauthor 
Patrick Woudt$^{8}$, Elmar K\"{o}rding$^{7}$, Rudolf Le Poole$^{10}$ \& Danielle Pieterse$^{7}$. \\
$^{1}$Physics \& Astronomy, The University of Southampton, SO17 1BJ, UK\\
$^{2}$ South African Astronomical Observatory, P.O Box 9, Observatory, 7935, Cape Town, South Africa\\
$^{3}$Department of Astronomy and Astrophysics, The Pennsylvania State University, 525 Davey Lab, University Park, PA 16802, USA\\
$^{4}$ Department of Astronomy, University of Cape Town, Private Bag X3, Rondebosch 7701, South Africa\\
$^{5}$University of Leicester, Astrophysics Group, School of Physics \& Astronomy, University Road, Leicester LE1 7RH, UK\\
$^{6}$Astronomical Observatory, University of Warsaw, Al. Ujazdowskie 4, 00-478 Warszawa, Poland\\
$^{7}$ Department of Astrophysics/IMAPP, Radboud University, P.O. 9010,
6500 GL, Nijmegen, The Netherlands \\
$^{8}$ Department of Astronomy \& Inter-University Institute for Data
Intensive Astronomy, University of Cape Town, Private Bag X3, 7701
Rondebosch, South Africa \\
$^{9}$ IAU-Office For Astronomy for Development, P.O. Box 9, 7935
Observatory, South Africa \\
$^{10}$ Leiden Observatory, Leiden University, P.O. Box 9513, NL-2300 RA
Leiden, The Netherlands \\
}

\date{Accepted XXX. Received YYY; in original form ZZZ}

\pubyear{2021}

\begin{document}
\label{firstpage}
\pagerange{\pageref{firstpage}--\pageref{lastpage}}
\maketitle

\begin{abstract}
SXP 15.6 ~is a recently established Be star X-ray binary system (\bexrb) in the Small Magellanic Cloud (SMC). Like many such systems the variable X-ray emission is driven by the underlying behaviour of the mass donor Be star. It is shown here that the neutron star in this system is exceptionally close to spin equilibrium averaged over several years, with the angular momentum gain from mass transfer being almost exactly balanced by radiative losses. This makes SXP 15.6 exceptional compared to all other known members of its class in the SMC, all of whom exhibit much higher spin period changes. In this paper we report on X-ray observations of the brightest known outburst from this system. These observations are supported by contemporaneous optical and radio observations, as well as several years of historical data.
\end{abstract}




\section{Introduction}

\color{black}
\bexrbs\ are a large sub-group of the well-established category of High Mass X-ray Binaries (HMXB) characterised by being a binary system consisting of a massive mass donor star, normally an OBe type, and an accreting compact object, a neutron star (though there is one known system, MWC 656, where the accretor is a black hole \citep{2014casares}). The Small Magellanic Cloud (SMC) has been known for quite a while now to contain the largest known collection of \bexrbs\ - see, for example, \cite{ck2015, hs2016}. Despite the many observational studies it remains clear that the complex interactions between the two stars continues to produce unexpected surprises. In particular, the unpredictable behaviour of the mass donor OB-type star is major driver in the observed characteristics of such systems, and as a direct result of the rate of mass transfer onto the neutron star systems long-term spin up or spin down changes are observed \citep{klus2014} . 

\color{black}

It is rare to find a system that approaches long-term equilibrium and an essentially zero spin period change.
The source that is the subject of this paper, SXP 15.6, could be such a system and was identified as a \bexrb ~ by \cite{vasil2017}. The optical counterpart \m2002 ~ is proposed to have a similar spectral type by several authors :  O9.5Ve by \cite{Evans2004}, O9IIIe by \cite{lamb2016} and B0IV-Ve by \cite{mcbride2017}.

Reported here are multiwaveband observations covering the X-ray and optical bands over several years showing the pattern of changes seen in this system. There are three broad occasions when SXP 15.6 was detected by the SMC X-ray survey project S-CUBED \citep{kennea2018} for a period of several hundred days, with the most recent detection (November 2021) being the brightest so far seen. In particular, it is noted that the pulse period change over several years is shown to be extremely small for such a \bexrb ~system.
 
 \color{black}
\section{Observations}

\subsection{X-rays - S-CUBED}

SXP 15.6 was detected by the S-CUBED survey \citep{kennea2018}, a shallow weekly X-ray survey of the optical extent of the SMC by the Swift X-ray Telescope (XRT; \citealt{burrows05}). Individual exposures in the S-CUBED survey are typically 60s long, and occur weekly, although interruptions can occur due to scheduling constraints. Full details of the X-ray reduction and manner in which sources are identified and their fluxes quantified can be found in \cite{kennea2018}.

\begin{figure*}
	\includegraphics[width=16cm,angle=-00]{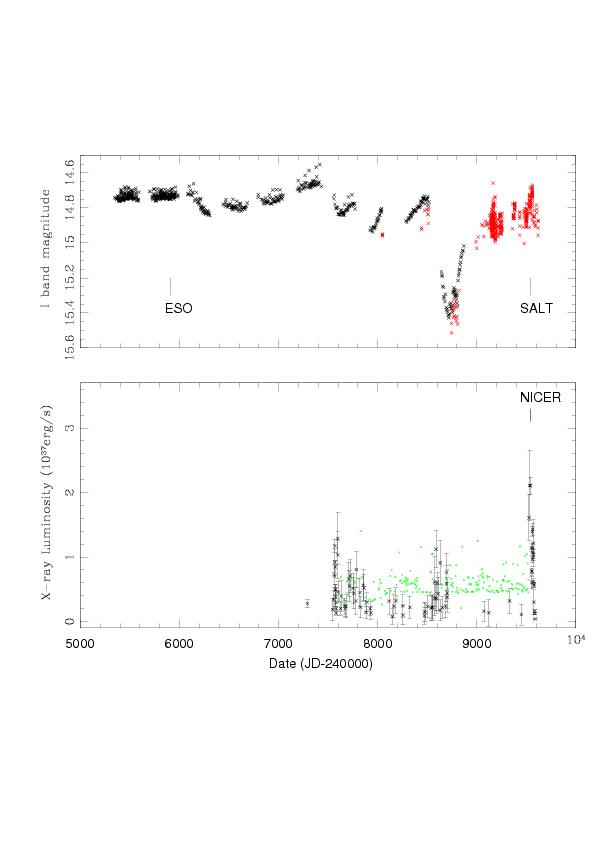}
    \caption{Long term S-CUBED X-ray (lower panel) \& I-band (upper panel) detections. The I band is a composite of OGLE IV I-band data (in black) and MeerLICHT i-band observations (in red). Note that MeerLICHT values have been adjusted for the slight differences in band passes - see the text. The green points in the lower panel indicate 90\% X-ray upper limits from non-detections. \color{black}The dates of the ESO, SALT and NICER observations are indicated.\color{black}}
    \label{fig:longXO}
\end{figure*}

\begin{figure}

	\includegraphics[width=8cm,angle=-00]{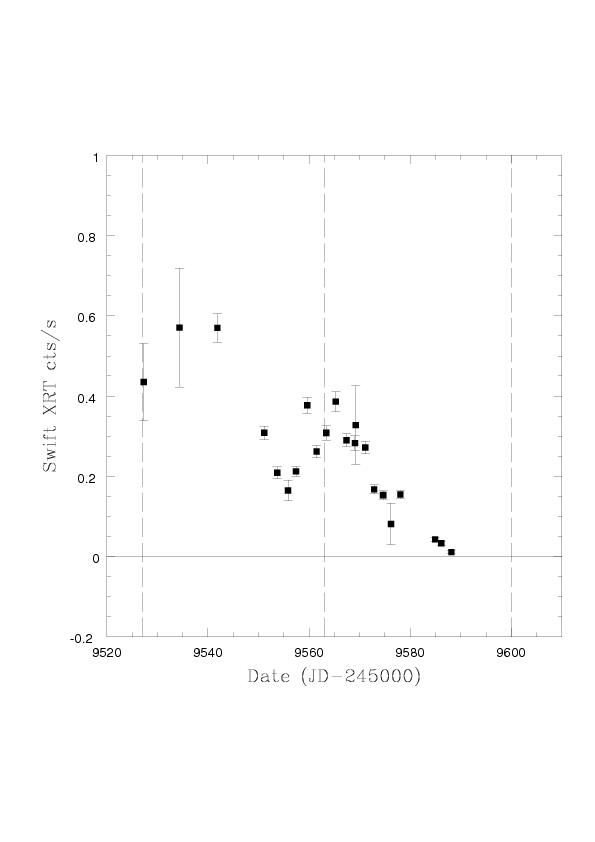}
    \caption{Swift XRT data during the current outburst (November 2021). The vertical dashed lines indicate the predicted time of optical outbursts given by Equation \ref{eq:1}. }
    \label{fig:swift_lc}
\end{figure}

SXP 15.6 was first detected by S-CUBED observation taken on MJD 57547 (2016 June 6) and reported by \cite{evans2016}. It has been seen many times since then - see Figure ~\ref{fig:longXO}. There have been a total of 280 S-CUBED observations over the $\sim$6 year period (91 detections and 189 upper limits), the results are shown in Figure ~\ref{fig:longXO}. It is noticeable that there have been three distinct periods of X-ray activity over these $\sim$5 years, with the most recent one being the X-ray brightest \citep{coe2021}. A detailed plot of the most recent outburst is shown in Figure ~\ref{fig:swift_lc} which shows that, at its peak, the XRT count rate was $ 0.56 \pm 0.05$ cts/s. 
 Using a standard SMC distance of 62~kpc \citep{scowcroft2016} and correcting for absorption fixed at the value derived from \cite{Willingale2013}, this corresponds to a peak 0.3-10~keV luminosity of ${(1.8 \pm 0.2)} \times 10^{37}$~erg$\cdot {\rm s}^{-1}$.

\subsection{X-rays - NICER}
\color{black}

Follow up observations were carried out by the Neutron Star Interior Composition Explorer (NICER) \citep{NICER}. A NICER 7.3ks observation for SXP 15.6 was obtained beginning 02:04 UT on Nov 20, 2021 and the results first reported by \cite{coe2021} who revealed the presence of a significant detection of the pulsar period at 15.64034(1)s based on the first NICER observation. This same pulsar period was also weakly detected in the Swift/XRT data.

In order to more accurately measure the spin period and evolution of the pulsar, we have analysed the first 7 days of NICER observations, in which the source was both bright and densely observed, with 8 observations being taken between between 2021 November 20 and 2021 November 27, representing a total of 31.6ks of observations during that time period. 

These data were processed using the standard HEAsoft 6.30 tools\footnote{\cite{HEAsoft}}, with barycentric corrections applied using the \texttt{barycorr} tool. A photon Time of Arrival (TOA) analysis was performed, in order to accurately measure the pulsar frequency and derivatives over this period of intense monitoring. This TOA analysis utilized the NICERsoft\footnote{\texttt{https://github.com/paulray/NICERsoft}} tools to perform the required pulse profile modelling, splitting of event data and TOA analysis. These resultant TOAs were then fit to find the pulsar frequency and its derivatives utilizing the PINT pulsar timing software package \cite{PINT}, specially utilizing the \texttt{pintk} user interface tool to perform the TOA fit.

Utilizing these tools, it was possible to find a phase locked period solution to the first 7 days of NICER observations. The resultant frequency fit is given in Table~\ref{tab:TOA}, and gives a best fit pulsar period of $15.639743 \pm 0.000022$~s. It is noteworthy that this pulsar period is consistent within errors with the value of $15.6398 \pm 0.0009$~s reported in a 2016 observations by \cite{vasil2017}, suggesting no strong evolution of the spin period in the last 5 years. 

The NICER derived value of  $\dot f$, which equates to a pulsar spin-up $\dot P$ of $-1.60 \pm -0.20 \times 10^{-9}$ s~s$^{-1}$, or approximately $-0.051 \pm 0.06$ s~year$^{-1}$ suggests that there is additional spin-up of the pulsar occurring during the X-ray outburst due to accretion. As the implied period derivative between 2016 and 2021 observations, $\dot P = -0.12 \pm 2.0 \times 10^{-4}$~s~year$^{-1}$), is much smaller than this this instantaneous value,  this suggests that this spin-up is not occurring during periods of quiescence, and only during the short period of outburst.

\begin{table}
\caption{Pulsar timing parameters for SXP 15.6 based on the the first 7 days of NICER observations.}
\label{tab:TOA}
\begin{tabular}{ll}
\hline
Parameter & Value\\
\hline
Epoch (MJD)       &     59538.087\\
$f$ (Hz)           &        0.06393966(9) Hz\\
$\dot f$ (Hz/s)             &     $6.544 \pm 0.819 \times 10^{-12}$\\
$\ddot{f}$ (Hz/s$^2$)            &    $-1.382 \pm 0.292 \times 10^{-17}$\\
\hline
\end{tabular}
\end{table}

This lack of change in the pulse period over 5-6 years is discussed below.

\color{black}

\subsection{OGLE IV}

The OGLE project \citep{Udalski2015}  provides long term I-band photometry with a cadence of 1-3 days. The star \m2002 ~was observed continuously for over a decades until COVID-19 restrictions prevented any further observations after March 2020. It is identified in the OGLE catalogue as:\\
\\
OGLE IV (I band): SMC720.11.13342 \\
OGLE IV (V band):  SMC720.11.20699v \\

The I-band lightcurve produced from the OGLE IV observations is shown in Figure ~\ref{fig:longXO}.

\subsection{MeerLICHT}

\m2002 ~ was monitored with the MeerLICHT telescope using SDSS filters (\textit{u,g,r,i,z}) and a wider \textit{q-}band filter ($4400 - 7200$~\AA) at 60~s intergration time for each filter. MeerLICHT is a prototype of the Black GEM array \citep{2019NatAs...3.1160G} and is primarily built to provide simultaneous sky coverage as the MeerKAT radio telescope. The telescope comprises a 0.65~m primary mirror with a 110~Megapixel CCD, resulting in a 2.7~deg$^2$ field of view \citep{2016SPIE.9906E..64B}. The MeerLICHT images were processed with the BlackBOX pipeline \citep{2021ascl.soft05011V}, which carries out primary reductions that include bias subtraction, overscan corrections and flat-fielding. The subsequent steps performed by the pipeline include astrometric calibration, estimation of the Point Spread Function as a function of position and photometric calibration. 

To complement the OGLE coverage, and to fill the gap after OGLE IV ended, the MeerLICHT i-band data have been added to Figure ~\ref{fig:longXO}. However, because the band passes for Johnson I and Sloan i are slightly different the MeerLICHT results have been adjusted according to the transformations given by \cite{jordi2006}. As a result the MeerLICHT values were adjusted by an amount of -0.31 magnitudes for display in this figure.

\begin{figure*}

	\includegraphics[width=18cm,angle=-00]{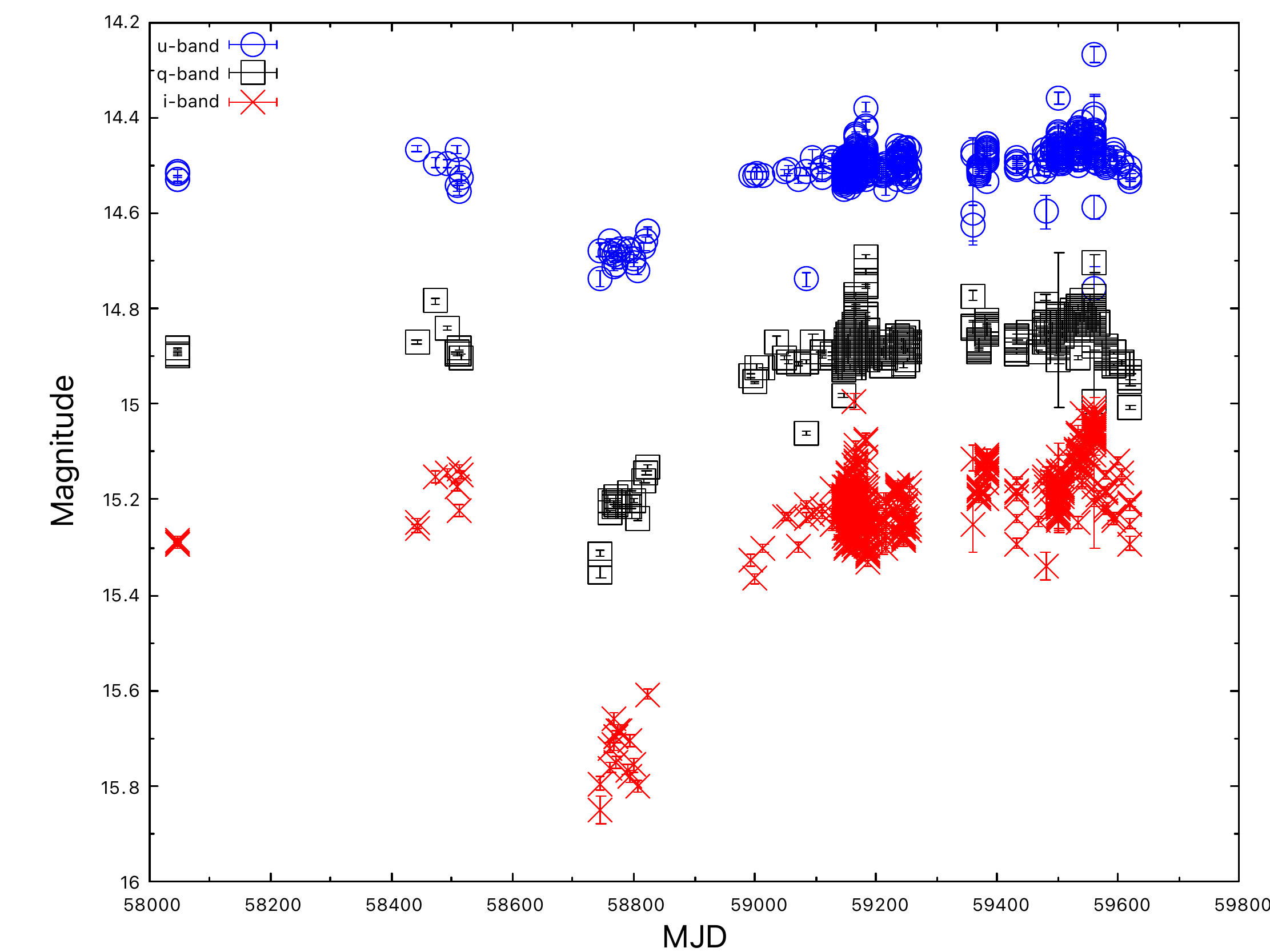}
    \caption{MeerLICHT $u-$ (blue open circles), $q-$ (black open squares) and $i-$band (red crosses) lightcurves. }
    \label{fig:ML_filter}
\end{figure*}

The unmodified lightcurves from the the MeerLICHT u, q and i band filters are shown in see Figure ~\ref{fig:ML_filter}. It is apparent from that figure that it is the red colour that changes most over time, indicative of the variable nature of the light from the circumstellar disc which is cooler, in general, than the star. This is quantified by comparing the size of the brightness decrease seen in all the filters around MJD 58800 compared to MJD 59200 - see Table \ref{tab:ML}. Note though SXP 15.6 was measured at the time of its faintest state in all filters, the general coverage in the g, r \& z bands is sparse so is not shown in Figure ~\ref{fig:ML_filter}.

\begin{table}
\caption{Table of maximum brightness decrease in SXP 15.6 seen by MeerLICHT.}
\label{tab:ML}
\begin{tabular}{ccc}
\hline
Waveband&Wavelength range \AA & $\Delta$m\\

\hline
u & 3400-4100 & 0.20$\pm$0.05 \\
g & 4100-5500 & 0.22$\pm$0.05 \\
r & 5600-6900 & 0.35$\pm$0.05 \\
q & 4400-7200 & 0.45$\pm$0.05 \\
i & 6900-8400 & 0.55$\pm$0.05 \\
z & 8400-10000 & 0.50$\pm$0.05 \\
\hline
\end{tabular}
\end{table}

\color{black}

\subsection{SALT \& ESO spectra.}

\m2002 ~was  observed with the ESO Faint Object Spectrograph and Camera v2. \citep[EFOSC2][]{Buzzoni1984}, mounted at the Nasmyth B focus of the 3.6m New Technology Telescope (NTT) at La Silla Observatory, Chile, on the night of 2011 December 11 (TJD 55907).The instrument was in longslit mode with a slit width of 1.5 arcsec and instrument binning 2$\times$2. Grism 20 was used to obtain spectra at ($\sim$6000 - 7000\AA{}) wavelengths. For Grism 20, this lead to a dispersion of $\sim$1\AA{}/pix and a resolution of $\sim$6\AA{}/fwhm. The spectra were reduced, extracted and calibrated using the standard \textsc{iraf}\footnote{Image Reduction and Analysis Facility: iraf.noao.edu} packages. 

\m2002 ~was also observed with the Southern African Large Telescope (SALT; \citealt{2006SPIE.6267E..0ZB}) using the Robert Stobie Spectrograph \citep{2003SPIE.4841.1463B,2003SPIE.4841.1634K} on 19 November 2021 (MJD59538) and 8 December 2021 (MJD59557). The PG2300 grating was used with an exposure time of 1200 seconds covering a wavelength range $6100 - 6900$~\AA. The SALT science pipeline \citep{2012ascl.soft07010C} was used to perform primary reductions, which include overscan corrections, bias subtraction, gain and amplifier cross-talk corrections. The remaining steps, comprising wavelength calibration, background subtraction and extraction of the one-dimensional spectrum were executed with \textsc{iraf}. 

\begin{figure}

	\includegraphics[width=8cm,angle=-00]{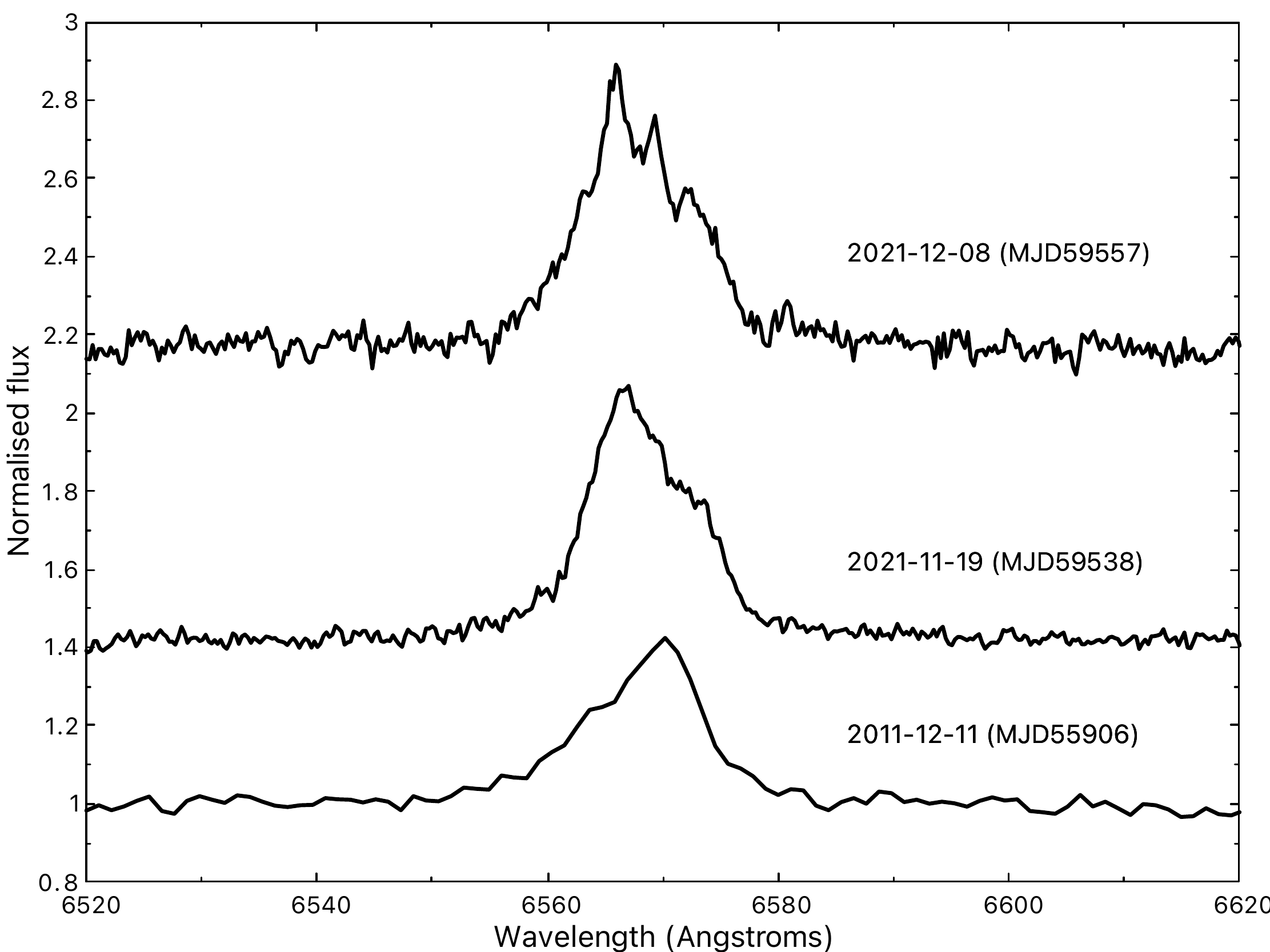}
    \caption{H alpha from ESO \& SALT- ITU. H$\alpha$ emission line (bottom is from ESO and top 2 are from SALT)}
    \label{fig:halpha}
\end{figure}

The SALT H$\alpha$ spectra compared to the historic ESO spectrum are shown in Figure ~\ref{fig:halpha}. The measured H$\alpha$ equivalent width values are MJD 55906 $-6.12 \pm0.31$\AA, ~MJD 59538 $-8.20 \pm0.30$\AA ~and MJD 59557 $-9.40 \pm0.47$\AA. This indicates a $\sim$50\% flux increase in this line between December 2011 and  November 2021. The most recent line profiles reveal multiple structural features which are only hinted at in the earliest spectrum. Presumably these features are related to changing structures in the circumstellar disc as it goes through a re-building phase and interacts gravitionally with the orbiting neutron star, thereby triggering the observed X-ray emission. 

\subsection {MeerKAT}
\begin{figure*}
	\includegraphics[width=16cm,angle=-00]{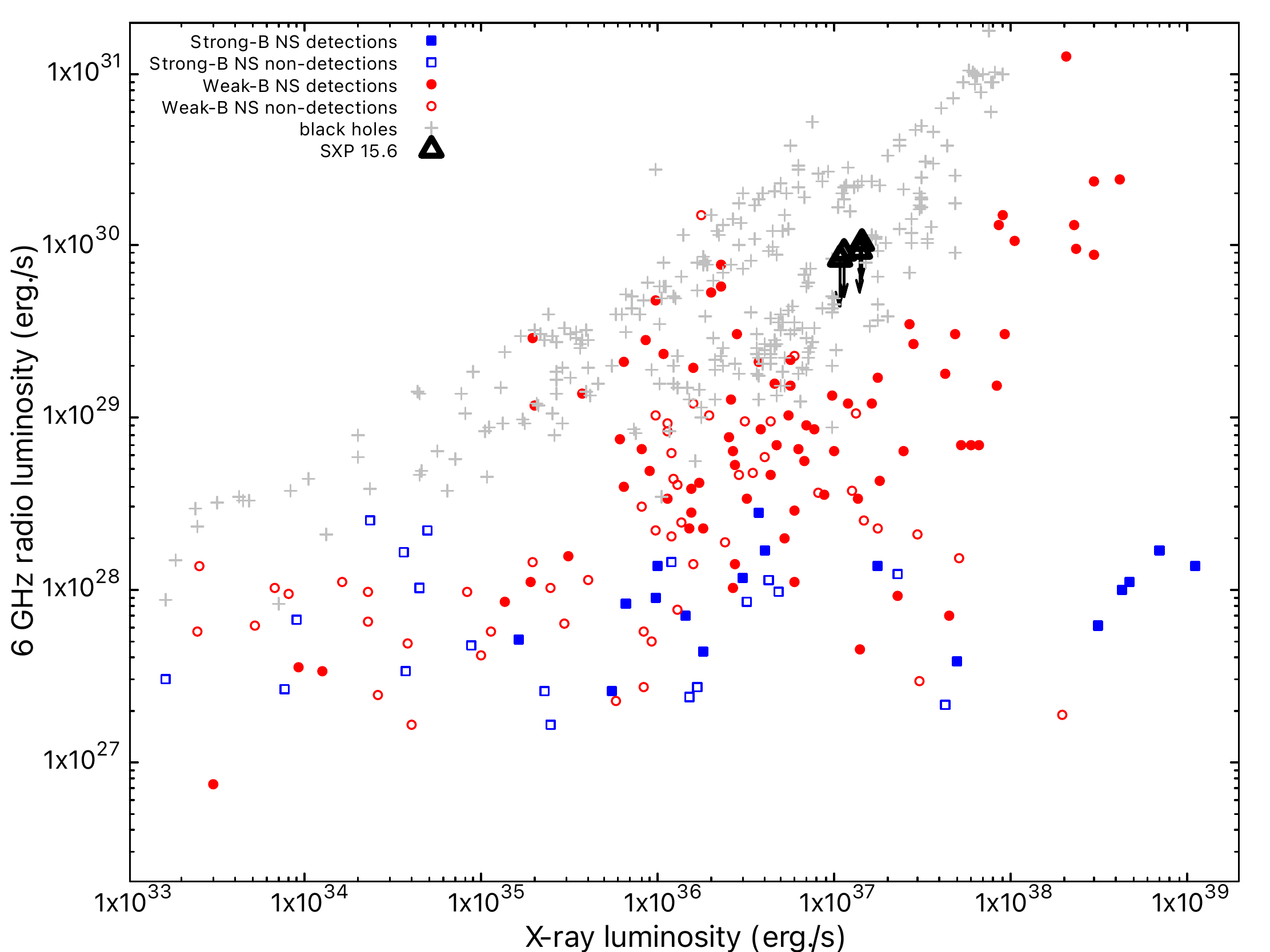}
    \caption{The radio/X-ray correlation for X-ray binaries. The measurements for SXP~15.6 are shown with the triangle symbols. Archival measurements of strongly-magnetised and weakly-magnetised neutron star systems are shown with the filled squares and circles, respectively. Radio upper limits of the strongly-magnetised and weakly-magnetised neutron star systems are shown with the open squares and circles, respectively. Archival black holes are shown in grey plus symbols. The archival data in this plot are taken from the X-ray binary database compiled by Bahramian et al. (2018) and van den Eijnden et al. (2021) (see their Fig. 1).}
    \label{fig:LRLX}
\end{figure*}

We observed SXP 15.6 with the MeerKAT radio array \citep{2009IEEEP..97.1522J} during the current outburst at a central frequency of 1.28 GHz for four epochs. The observations were done with a bandwidth of 856 MHz, with the correlator configured to deliver 4096 channels. Each observation consisted of 1 hour scans of the target and 2 minute scans of the phase calibrator J0252-7104. We used J0408-6545 as the primary calibrator, which was observed at the start and end of each observation for 5 minutes. We processed the data with the Oxkat \citep{2020ascl.soft09003H} reduction routines. Oxkat contains scripts that average the data to 1024 channels, applies standard bandpass and gain corrections, followed by flagging of the data. The target field was then imaged using \textsc{wsclean}, and then a process of self-calibration was executed using \textsc{cubical}. All the images resulted in a non-detection at the optical position of SXP 15.6 ($\sim$1~arcsec positional error). The 3$\sigma$ upper limits are given in Table~\ref{tab:MeerKAT}. In Fig.~\ref{fig:LRLX} we plot the 3$\sigma$ radio luminosity upper limits with the simultaneous X-ray luminosities. In this plot the MeerKAT flux densities were converted to 6GHz luminosities by assuming a flat spectrum and using the SMC distance of 62 kpc \citep{scowcroft2016}. For comparison, we include X-ray and radio luminosity measurements of X-ray binaries from \cite{arash_bahramian_2018_1252036} and \cite{2021MNRAS.507.3899V}.  

\begin{table}
	\centering
	\caption{MeerKAT (1.28~GHz) upper limits of SXP~15.6 at the 3$\sigma$ level}
    \setlength\tabcolsep{2pt}
	\begin{tabular}{ccc} 
		\hline\hline
		MJD & Radio flux density ($\mu$Jy) & Luminosity (erg./s)  \\
		\hline
		59559	&	$<$33.9 & $< 2.0\times 10^{29}$\\
		59563   &   $<$32.7 & $< 1.9\times 10^{29}$\\
        59565	&	$<$37.5 & $< 2.2\times 10^{29}$\\
        59566	&	$<$30.6 & $< 1.8\times 10^{29}$\\        
		\hline
	\end{tabular}
	\label{tab:MeerKAT}
\end{table}

\section{Discussion}

\subsection{Long term spin period changes}

\begin{figure}

	\includegraphics[width=8cm,angle=-00]{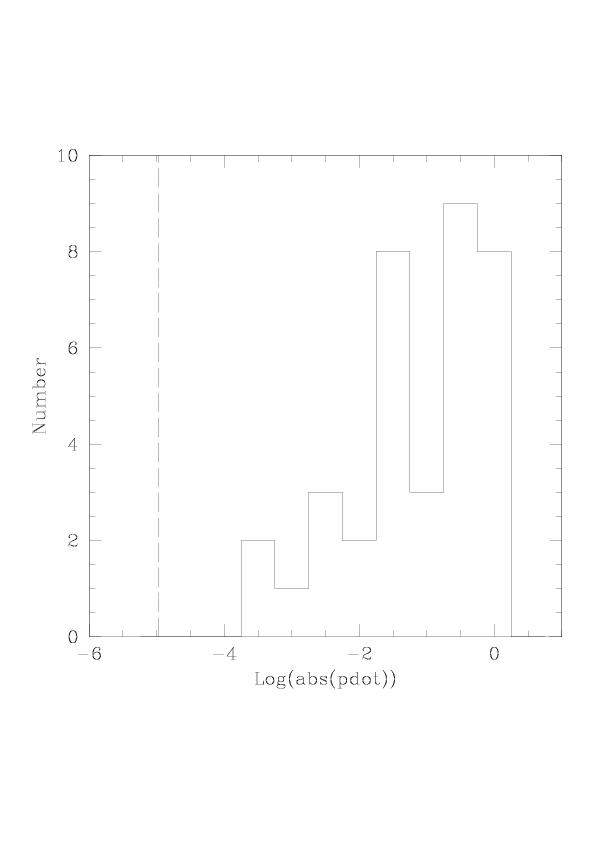}
    \caption {Histogram of log of the absolute value of $P_{dot}$ ($P_{dot}$ in units of s/yr) for 42 SXP systems averaged over 10 years \citep{klus2014}. The vertical dashed line shows the position of SXP 15.6 from this work.}
    \label{fig:pdot}
\end{figure}

\color{black}
\cite{klus2014} published a study of the spin period changes, $P_{dot}$, in 42 \bexrb ~systems in the SMC. From their results it is possible to determine the average $P_{dot}$ over more than a decade for these systems - see  Figure ~\ref{fig:pdot}. Comparing the NICER spin period of SXP 15.6 with that of the Chandra measurement in July 2016 \citep{vasil2017}, it is possible to determine that the average spin period change over the $\sim$6 years is $-2.91 \times 10^{-8}$ s/day or $-1.06 \times 10^{-5}$ s/year. This is an extremely small absolute value of $P_{dot}$, by far the smallest of all the 43 systems so far measured in the SMC. This exceptionally low spin period change is illustrated in Figure ~\ref{fig:pdot}.
\color{black}
Of all the 43 \bexrb ~sources currently known in the SMC with spin periods, SXP 15.6 is by far the closest to exhibiting spin equilibrium, at least in the last 6 years. \cite{klus2014} review models for accretion on to neutron stars and their Equation 18 (based on the work by \cite{davidson1973}) permits the determination of the neutron star magnetic field under such equilibrium circumstances:

\begin{equation}
B\approx1.8 \times 10^{13}  R^{-3} (\frac{M}{M_{\odot}})^{5/6} (\dot{M})^{0.5} (\frac{P_{spin}}{100})^{7/6} ~\textrm{~G}
\label{eq:klus}
\end{equation}
 where $R$ is the neutron star radius in units of $10^{6}$cm, $M$ is the mass of the neutron star, $\dot{M}$ is the mass accretion rate in units of gm/s and $P_{spin}$ is the spin period in seconds.
 
 Normal values are assumed here for the neutron star mass (1.4$M_{\odot}$) and radius (10 km) The spin period for SXP 15.6 is, of course, 15.6s. To evaluate $\dot{M}$ it was first necessary to find the average X-ray luminosity from the 280 S-CUBED observations over the $\sim$6 year period (91 detections and 189 upper limits). That was found to be $1.7 \times 10^{36}$ erg/s  - non-detections were assigned an arbitrarily low luminosity value of $10^{33}$ erg/s. Using that average luminosity and assuming a mass-to-energy conversion efficiency of 10\% , gives $\dot{M} = 2.1 \times 10^{16}$ gm/s. Putting that number back into Equation \ref{eq:klus} results in a predicted magnetic field of $3.7 \times 10^{12}$ G for the neutron star in SXP 15.6.

\subsection {Comparing X-ray with optical lightcurves}

Figure ~\ref{fig:longXO} shows the totality of our optical and X-ray data on SXP 15.6. The OGLE IV I-band data have been supplemented with MeerLICHT i-band. In the figure the MeerLICHT data have been adjusted by 0.36 magnitudes so that they agree with OGLE IV where they overlap. This is necessary as the two filters used (I and i) have slightly different bandwidth responses. 

\begin{figure}

	\includegraphics[width=8cm,angle=-00]{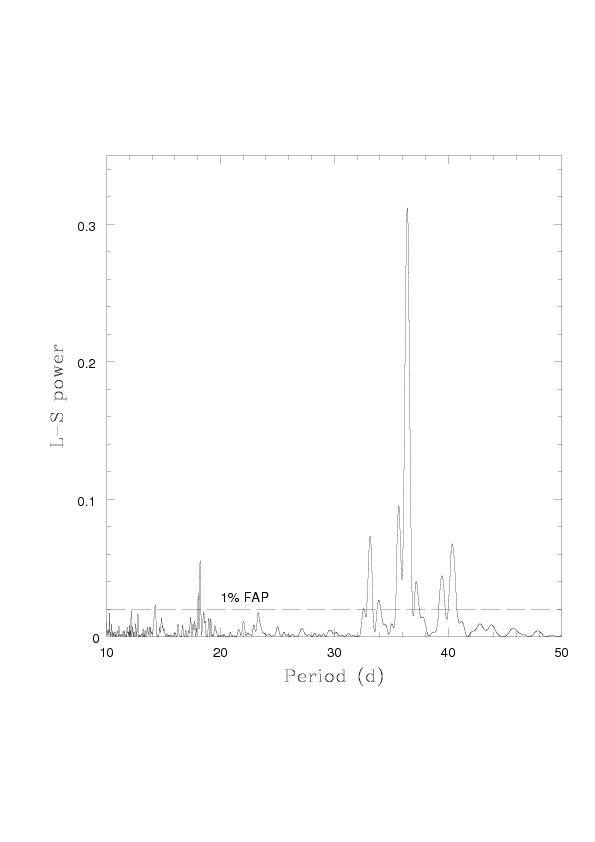}
    \caption{Generalised Lomb-Scargle power spectrum from the OGLE IV data. The peak is at 36.411 d.}
    \label{fig:lspower}
\end{figure}

A visual inspection of the OGLE IV data in this figure quickly reveals the existence of short optical spikes, predominantly obvious during the period 5346 -- 7772 TJD. Taking that interval of data, detrending it and then applying a Generalised Lomb-Scargle timing analysis \citep{gls2009} reveals a strong, clear peak in the power spectrum at a period of 36.411d - see Figure ~\ref{fig:lspower}.

The ephemeris for the time of the optical outbursts, $T_{opt}$, is here updated from the earlier measurements of \cite{mcbride2017} and is now given by  : \\
\begin{equation}
T_{opt} = 2455376.41 + N(36.411) ~\textrm{~JD}
\label{eq:1}
\end{equation}

\begin{figure}

	\includegraphics[width=8cm,angle=-00]{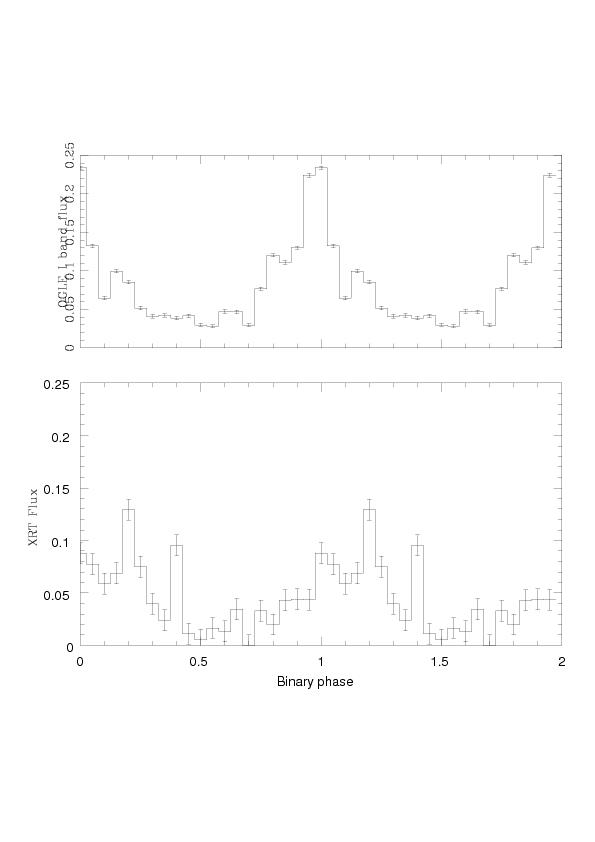}
    \caption{OGLE IV I-band data (upper panel) and S-CUBED X-ray data (lower panel) folded at the binary period given in Equation \ref{eq:1}}
    \label{fig:double}
\end{figure}

\
The S-CUBED X-ray flux is compared to the OGLE IV fluxes, both folded with the ephemeris given in Equation \ref{eq:1}, in Figure ~\ref{fig:double}. It is immediately clear that the X-ray emission is only weakly correlated with the OGLE modulation, both in width and peak position. Though the gravitational pull of the arriving neutron star is believed to extend the surface area of the optically thick circumstellar disk thereby increasing the I band emission, it seems to have little effect on the material accreting on to the neutron star and triggering X-ray emission. There is some suggestion that the period of X-ray maximum lags the optical peak by a phase of $\sim$0.2.

More generally, it can be seen from Figure ~\ref{fig:longXO} that the times when the source is most X-ray active is correlated with the source exhibiting its more normal bright state. The dip in the I-band around TJD 8800 of some 0.7 magnitudes coincides with the period when S-CUBED was failing to detect the system. In contrast, the most recent X-ray bright state around TJD 9500-9600 marks the recovery of the circumstellar disc, and the resultant availability of material for accretion. 

\color{black}

\subsection {OGLE and MeerLICHT colour-magnitude diagrams}

\begin{figure}

	\includegraphics[width=8cm,angle=-00]{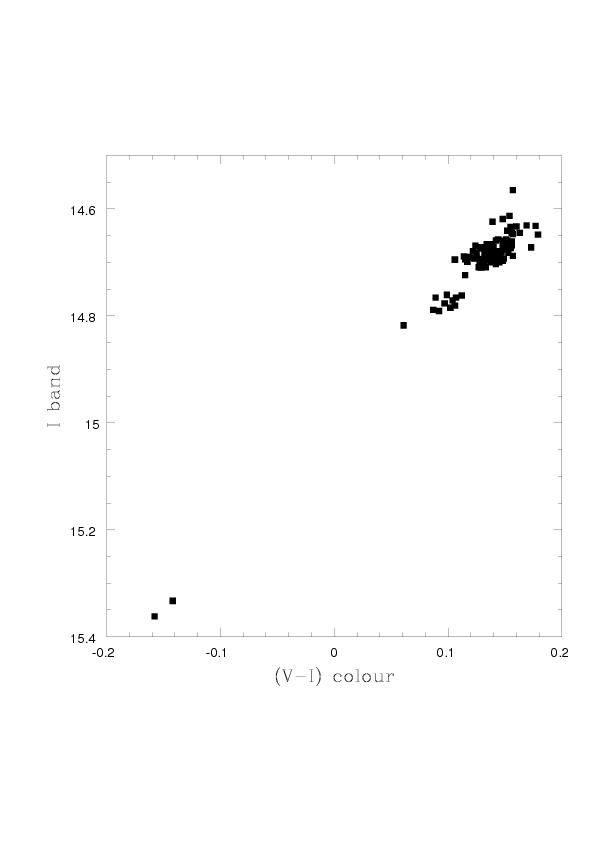}
    \caption{OGLE IV colour-magnitude diagram from data collected over $\sim$9 years.}
    \label{fig:ogle_cmd}
\end{figure}

The optical counterpart \m2002 ~ is proposed to be of spectral type O9.5Ve by \cite{Evans2004}, O9IIIe by \cite{lamb2016} and B0IV-Ve by \cite{mcbride2017}. The intrinsic (V-I) colour of a O9.5V - B0V is in the range -0.361 to -0.355  in \cite{pm2013}. The OGLE dust maps of the SMC \citep{skowron2021} enable the precise reddening correction to be made for such an object in the SMC and that is E(V-I)=0.067. Thus the predicted observed colours of \m2002 ~if it were a B-type star in this range with no circumstellar disk will be (V-I) = -0.294 to -0.288. From Figure ~\ref{fig:ogle_cmd} it can be seen that the observed colours are always much redder than this value, even during the epoch when the source was at its faintest for a few years. This strongly suggests the presence of a persistent circumstellar disc, adding further reddening to the observed colours. However, it is worth noting that the system has a clear pattern of being much bluer when fainter, strongly suggesting that all the variations seen in the I band magnitude are due to variations in this disc size. 

\begin{figure}

	\includegraphics[width=8cm,angle=-00]{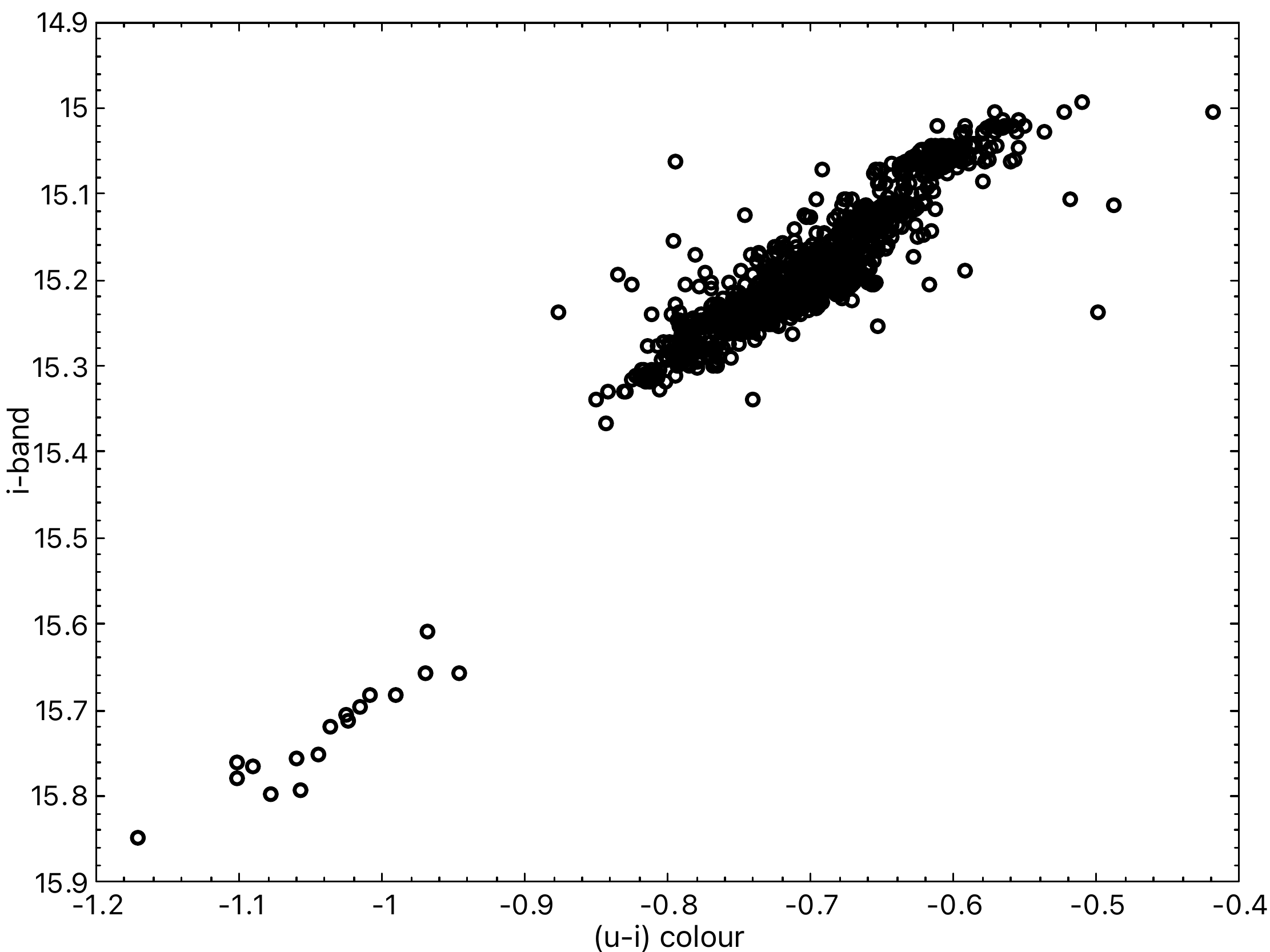}
    \caption{MeerLICHT CMD - data from just the recent outburst epoch.}
    \label{fig:meerlicht}
\end{figure}

Fig. \ref{fig:ogle_cmd} and Fig.~\ref{fig:meerlicht} reveal a strong correlation between the brightness and colour. This is an indicator of a low/intermediate disc inclination, since the growth results in overall excess brightness and reddening of the system as the outer parts of the disc are cooler than the inner parts. The inference of a low disc inclination is corroborated by the single-peak morphology of the H$\alpha$ emission line (Fig.~\ref{fig:halpha}).

\color{black}
\subsection{Circumstellar disc parameters}

To estimate the probable dimensions of the disk and neutron star orbit some assumptions are needed. 

The EW of the H$\alpha$ emission line may be used as a gauge to the size of the disc and was shown to be correlated to radii measurements from optical interferometry of nearby isolated Be stars \citep{2006ApJ...651L..53G}.
Taking the H$\alpha$ typical value during the recent outburst as -9$\pm1$\AA ~ this predicts an H$\alpha$ emitting disk of radius 130-280 $R_\odot$ or $(0.9-1.9) \times 10^{11}$m. 

 Also assuming, for simplicity, that the neutron star is in a circular orbit, then the period of 36 d permits an estimate of the orbital radius to be $9.0 \times10^{10}$ m. That is in good agreement with the estimate for the disc size and, broadly speaking, this is what Smooth Particle Hydrodynamic simulations of such systems predict \citep{okazaki2001,brown2019}. Specifically that the neutron star orbit is constraining further disk expansion beyond that point.

\subsection{Radio emission}

The authors \cite{2021MNRAS.507.3899V} presented a comprehensive study of radio observations of neutron star X-ray binaries. In their work, they demonstrate that strongly-magnetised accreting neutron stars ($B\geq 10^{10}$~G), such as those in \bexrbs, can be detected at radio frequencies, contrary to what was previously thought. The Galactic \bexrb ~systems A0535+262 and Swift J~0243.6+6124 were detected in the radio during enhanced accretion states that resulted in Type II X-ray outbursts \citep{2019MNRAS.483.4628V,2020ATel14193....1V}. The radio emission in these systems is proposed to be due to the launch of an accretion-powered jet. The non-detection of radio emission from SXP~15.6 during its high X-ray state ($L_X > 10^{37}$~erg./s) possibly indicates a weak or absent jet. However, because SXP~15.6 is the only extragalactic X-ray binary in Fig.~\ref{fig:LRLX}, it is only just possible to reach the needed sensitivity to detect the fluxes observed from previously reported galactic systems. It is expected that future radio telescopes in the southern hemisphere should soon make this goal much more achievable, and the large sample of \bexrb ~systems in the Magellanic Clouds will be important targets for further understanding of the radio emission from such HMXBs.

\section{Conclusions}

\bexrb ~systems have been known for a long time to show both spin up and spin down signatures, depending, it is believed, upon the average accretion rates of material on to the neutron star. However, the system that is the subject of this paper, SXP 15.6, is demonstrating the closest example yet known where spin equilibrium appears to prevail over the several years it has been studied. Such a system is rare and it provides a valuable example in understanding the panoply of accretion mechanisms that are proposed for these \bexrb ~systems.

\section*{Acknowledgements}

The OGLE project has received funding from the National Science Centre, Poland, grant MAESTRO 2014/14/A/ST9/00121 to AU. PAE acknowledges UKSA support. JAK acknowledges support from NASA grant NAS5-00136. This work made use of data supplied by the UK Swift Science Data Centre at the University of Leicester.

PJG is supported by NRF SARChI grant 111692.

The MeerLICHT telescope is designed, built and operated by a consortium
consisting of Radboud University, the University of Cape Town, the
South African Astronomical Observatory, the University of Oxford, the
University of Manchester and the University of Amsterdam, with support
from the South African Radio Astronomy Observatory.

The MeerKAT telescope is operated by the South African Radio Astronomy Observatory, which is a facility of the National Research Foundation, an agency of the Department of Science and Innovation.

IMM, DAHB, VM and PW are supported by the South African NRF. 

Some of
the observations reported in this paper were obtained with the Southern African Large Telescope (SALT), as part of the Large Science
Programme on transients 2018-2-LSP-001 (PI: Buckley)

NICER is a 0.2-12 keV X-ray telescope operating on the International Space Station. The NICER mission and portions of the NICER science team activities are funded by NASA.

\section*{Data Availability}

 All X-ray data are freely available from the NASA Swift and NICER archives. The OGLE optical data in this article will be shared on any reasonable request to Andrzej Udalski of the OGLE project. Requests to access the MeerLICHT data should be addressed to Paul Groot.



\bibliographystyle{mnras}
\bibliography{references} 



\bsp	
\label{lastpage}
\end{document}